# Embedding Economic Incentives in Social Networks Shape the Diffusion of Digital Technological Innovation

Zhe LI[a], Tian-fang ZHAO[c], Hong-jun ZHU[ab,*]

[a]School of Journalism and Communication, University of Chinese Academy of Social Sciences, Beijing, China
[b]Institute of Journalism and Communication, Chinese Academy of Social Sciences, Beijing, China
[c]School of Journalism and Communication, Jinan University, Guangzhou, China

**Abstract**

The digital innovation accompanied by explicit economic incentives have fundamentally changed the process of innovation diffusion. As a representative of digital innovation, NFTs provide a decentralized and secure way to authenticate and trade digital assets, offering the potential for new revenue streams in the digital space. However, current researches about NFTs mainly focus on their transaction networks and community culture, leaving the interplay among diffusion dynamics, economic dynamics, and social constraints on Twitter. By collecting and analyzing NFTs-related tweet dataset, the motivations of retweeters, the information mechanisms behind emojis, and the networked-based diffusion dynamics is systematically investigated. Results indicate that Retweeting is fueled by Freemint and trading information, with the higher economic incentives as a major motivation and some potential organizational tendencies. The diffusion of NFT is primarily driven by a 'Ringed-layered' information mechanism involving individual promoters and speculators. Both the frequency and presentation of content contribute positively to the growth of the retweet network. This study contributes to the innovation diffusion theory with economic incentives embedded.

## 1. Introduction

In the ongoing landscape of digitalization, diffusion process plays a crucial role in determining the contribution of digital innovation to economic growth and social welfare (Fagerberg et al., 2005). Social networks exert influence on the digital innovation through the extensive connectivity and network effects (Van Dijck, J., 2013). The integration of economic behavior into social networks highlights how network structures impact the economic logic. Against this backdrop, the diffusion of digital innovation has become a focus for many researchers in communication, economics, marketing, and other fields. However, there is still few empirical evidence about how digital innovation diffuses through social networks and affects the embeddedness of economic behavior in this process.

As a representative and popular digital innovation, Non-Fungible Tokens (NFTs) rely almost entirely on social networks for issuance and transactions, generating significant economic impact. Until June 25, 2023, approximately 8.61 billion NFTs have been issued on 14 blockchains, including Ethereum, Solana, and Polygon, with around 120 million cumulative addresses and a total of 21.13 billion transactions, creating a market value of approximately $12.582 billion (NFTSCAN,2023). NFTs is an economic product of various digital technologies such as network science, blockchain, and artificial intelligence, whose global popularity presents a typical innovation diffusion trend involving economic incentives.

---

[1] Authors' addresses: Zhe LI:lizhe1@ucass.edu.cn; Tian-fang ZHAO:tfzhao@jnu.edu.cn; Hongjun ZHU[*]: zhuhongjun@ucass.edu.cn

This paper focuses on the overarching question of how economical digital innovations diffuse in social media platforms, and delves into the diffusion mechanism of NFTs on Twitter (renamed as X in 2023). Leveraging an embedded perspective, the motivations of economic behavior are emphasized to provide a fully explanatory framework for the network structure and social content of NFT diffusion on Twitter. Network analysis and text mining methods are adopted to reveal the diffusion dynamics and popular topics of NFT. A gear-like 'Ringed-layered' pattern composes of multiple star networks is found. Speculators, individual promoters, and market service entities are the main participants. Retweeting is propelled by Freemint and transaction information, with the pursuit of higher economic incentives being a major motive for this behavior. Additionally, mutual assistance and sharing industry information are also purposes for retweeting. The behavior of some retweeters demonstrates potential organizational tendencies. The activity of posters and the use of emojis in content both positively contribute to the growth of the retweet network.

The main contribution of this paper is to supplement the landscape of the diffusion of digital innovation and to show the interaction between economic logic and network structure. As a digital innovation without clear use cases and inconspicuous impacts on real life but possessing explicit economic value, NFTs serve as a blueprint for the diffusion of innovation in the future digital society. The 'Ringed-layered' pattern observed in the diffusion of NFT provides a new network structure illustration for digital marketing practices conducted in social networks. The core of this pattern highlights the significance of broadcasting spread in marketing within social networks.

To begin, the related work on the diffusion of innovation and embeddedness of economic behaviors in social networks is reviewed. Subsequent is the dataset, network analysis and text mining methods, along with the results. Following that, research findings are summarized. Finally, the limitations and provide future research prospects are discussed.

## 2. Literature Review

### *2.1 Innovation Diffusion in Social Networks: Structure determines the information mechanism*

The diffusion of innovation is a broad and mature research field (Guttentag & Smith, 2022). Diffusion networks is one of the major types in this field, focusing on examining connectivity patterns among members within a system (Rogers, 2003, p.98). Recent relevant research has shifted towards mathematical modeling to predict overall adoption trends, aiming to further emphasize the constraining role of network structures. A crucial question emerges regarding which network structural features influence and determine innovation diffusion (Peres et al., 2010).

Studies generally address this question on three levels. Firstly, at the macro level, attention is given to the impact of global structural features such as average degree, degree distribution, and clustering. Higher average degrees are believed to lead to faster diffusion (Mukherjee, 2014), while degree distribution provides insights into assessing asymmetric influence and diffusion time (Dover et al., 2012). Clustering may either accelerate diffusion through more efficient communication among homogeneous individuals or limit diffusion to the majority of the network by causing redundancy in the network structure (Rogers, 2003, p.326) when assumed to be equivalent to homogeneity (Bohlmann et al., 2010).

Secondly, at the intermediate level, the focus is on the impact of the strength of relationships on the diffusion. Research has extensively discussed the influence of weak and strong ties, generally acknowledging the 'strength of weak ties'. However, the relative impact cannot be uniformly generalized due to various factors such as individual network size, number of weak ties, advertising campaigns, characteristics of innovation, and strength of network externalities (Goldenberg et al., 2001).

The third aspect involves a micro-level exploration of the impact of hubs and connectors on the diffusion. Opinions on the impact of hubs differ, with some believing that hubs can facilitate broadcast, thus accelerating innovation diffusion (Pei et al., 2014; Liu et al., 2018). Others posit that the role of hubs is limited, as they only influence a finite number of neighbors (Hinz et al., 2011). Some studies even take a negative view, emphasizing that hubs disproportionately triggering large-scale cascades are exceptions rather than the rule. In contrast, in most cases, hubs are only slightly more important than ordinary individuals (Watts & Dodds, 2007). Research on connectors generally aligns in acknowledging their impact, considering them as critical nodes bridging structural holes in the network (Hinz et al., 2011), which can generate information advantages and expedite diffusion by establishing connections with different communities (van den Bulte & Wuyts, 2007).

Based on structural functionalism, these studies assert that the structural features of a network influence the diffusion process of innovation because they fundamentally construct the information mechanism determining 'who knows what and when' (Aral, 2021,p.77). Information constitutes the first stage of the innovation decision-making process (Rogers,2003,p.22), and the information mechanism is akin to determining the overall pattern of innovation diffusion in the networks (Krippner & Alvarez, 2007).

However, current research on NFTs has yet to provide a detailed explanation of the information mechanism based on network structure on the diffusion. Existing studies principally focus on analyzing transaction mechanisms based on network structural features, emphasizing the modularity, non-structural autonomy, and homogenization of transaction networks (Nadini et al., 2021). Considering that information is more advanced in the diffusion stage, it is essential to understand the information mechanism involved for a comprehensive analysis. In light of this, the paper poses the first research question:

**RQ1.** What structural features does the retweet network on Twitter for NFT-related topics exhibit, and what information mechanisms does it indicate?

*2.2 Economic Behavior Embedded in Social Networks: NFT Traits and Behavioral Motivation*

The research on the diffusion of innovation in social networks exhibits significant inadequacies in two aspects. First, the universality pursued by mathematical models partially squeezes out considerations for the characteristics of innovations, thus giving the impression that diffusion is based on a universal situation. Second, the emphasis on the objective network structure often overlooks how individuals are 'happen' connected in the network, where individual access to the network seems more like a mechanical process rather than a result of various motivational choices (Granovetter, 1978). Such diffusion models unsurprisingly fail to capture the complexity of real-world diffusion mechanisms (Kiesling et al., 2012).

Discussing the diffusion of NFTs in social networks requires a clear understanding of the key features. NFT is a smart contract on the blockchain representing digital ownership of physical assets, virtual assets, or other rights certificates (Ethereum Improvement Proposals, 2018; Chandra, 2022). The value of NFTs is mainly confirmed through the financial market, and the technical features of NFTs such as uniqueness (Sharma et al., 2022), indivisibility (Chandra, 2022), verifiability (Guadamuz, 2021), and interoperability (Moreaux & Mitrea, 2023) further paradoxically highlight the peer-to-peer resale financial opportunities. It is reasonable to believe that financial value is a crucial, undeniable feature of NFTs, setting them apart from objects traditionally examined in innovation diffusion theory. On the one hand, financial value, unlike external incentives such as fiscal subsidies (Münzel et al., 2019 & Simpson & Clifton, 2017), is endogenous to the innovative object. On the other hand, uncertain financial gains are generated through networked exchanges, not solely dependent on the innovation itself. Therefore, actions related to the creation, purchase, trade, and collection of NFTs should be viewed as economic behaviors.

Based on the unique financial features of NFTs, the theory of embeddedness from economic sociology provides a valuable supplement for understanding the diffusion of NFTs in social networks. The basic proposition of embeddedness theory is that economic behavior is embedded in social networks (Granovetter, 1985), aiming to avoid the atomistic tendencies implicit in over-and undersocialized conception of human actions by emphasizing that actors' economic behavior is purposefully embedded in specific, enduring social relationships (Granovetter, 1985). Early research on embeddedness focused on how structural embeddedness, especially the quality of relationships and positions in networks (Uzzi, 1999), influences economic behaviors such as pricing (Uzzi & Lancaster, 2004) and cooperation (Polidoro & Mitchell, 2011). Recent research has shifted focus to the causal role of network relationships on the analytical level, emphasizing relational embeddedness, cognitive embeddedness, and cultural embeddedness as consequences (Simsek et al., 2003; Dequech, 2003), yet still neglecting antecedents and the duality between antecedents and consequences.

Now, the primary question to address is what triggers behavior, not just explaining the form of behavior and its effects. To achieve this, an emerging trend is to support the embeddedness perspective by downward extending more explicit behavioral theories (Krippner & Alvarez, 2007). These studies assert that the embeddedness of economic behavior is formulated and shaped in the process of action (Beckert, 2003). Structure does not determine behavior, and purposeful interaction is central to understanding behavior (Crossley, 2010,p.142). These all emerge from the structure.

To understand how the structure is formed, that is, how the edges in the network are connected, we need to go back to the analysis of the actors' behavior in the context. While some studies have discussed the purposes of specific economic behavior in small-scale social networks through interviews and questionnaires (Cruz et al., 2013; Oh, 2007), there is limited analysis of the purposes of economic behavior in large-scale social networks. There is still little known about why people participate in the diffusion of NFTs in social networks. In light of this, this paper raises a second research question:

**RQ2.** What are the purposes behind retweeting NFT-related topics on Twitter?

## 3. Data

This study is grounded in data sourced from Twitter, a prominent global media platform that has positioned itself as a pivotal space for NFT issuance. Conducted as a cross-sectional analysis, the dataset contained content tagged with #NFT and #NFTCommunity from February 16, 2023, to March 18, 2023. There are two reasons to select this time range. Firstly, it represents a relatively stable period following the rapid development of NFTs, where the various types, issuance methods, and transaction modes have matured, and no novelty has emerged, resulting in the normalization of user activities. Secondly, this period has benefited from the incentive measures introduced by the new trading market, Blur. Discussions, issuance, and transactions of NFTs have recently peaked, creating an optimal environment for observing the diffusion of NFTs. The choice of these two hashtags is rooted in their prevalence as the most frequently used tags for relevant content during this specific time range.

The initial dataset consisted of 4,752,763 records featuring 10 fields: Date, UserID, Handle, Name, Text, URL, Platform, Type, Retweet count, and Favorites count. After necessary data cleaning procedures, including the removal of duplicates, consistency checks, handling missing values, and preliminary analysis, the dataset is finally refined to 4,369,750 data records. Extracting nodes and edges and further removing duplicate edges, a final retweet network comprising 911,376 nodes and 2,490,481 edges is obtained.

## 4. Methodology

The information mechanism of the diffusion of NFT on Twitter is concerned with the individual information transmission relationships in the network, as well as the mutual patterns and regularities of these relationships. Employing network analysis tailored for social structures, the retweet network of relevant topics is examined. Gephi is utilized to analyze the global features, community clustering, and key node features, focusing on average degree, degree distribution, and bidirectional edges. Modularity analysis utilizes the Louvain community detection algorithm (Vincent D Blondel et al., 2008). As this algorithm is stochastic, the mode is taken to determine the number of communities after running 10 times.

The analysis of key node features is conducted through manual labeling after selecting nodes from various positions and in-degrees within each community. Specifically, there are 107, 57, 73, 61, 42, 69 nodes are sampled from communities 1 to 6. For each node, the 'Handle' field is used to retrieve the profile and posts on Twitter. Following the classification by Wilson et al. (2022), nodes are categorized into 9 roles: projects, individual creators/artists, media/community and other market services, technical services, individual promoters/influencers, ordinal consumers, qualified consumers, and speculators. There is a certain level of ambiguity between qualified consumers, ordinary consumers, and speculators. Qualified consumers explicitly state themselves as OG, degen, Legends, etc., and typically use well-known NFT projects as avatars, and may also list their personal pages in trading markets. Ordinary consumers identify themselves as holders, lovers, or those interested in NFTs, and may also post their professions or other life-related content in their profiles. Speculators typically lack profiles, with all content consisting of freemint retweets.

To understand the purposes for retweeting, this study utilizes text mining to compare content differences between high-retweet nodes and low-retweet nodes. The goal is to discern which topics and representations of content are more likely to be retweeted and to explore the underlying purposes revealed in the discourse. Sorting non-zero in-degrees in descending order, selecting the top 25% nodes (16,838 nodes) as the Top group, and the bottom 25% nodes (50,512 nodes) as the Bottom group. Nodes are matched with original content, resulting in 255,266 tweets for the Top group and 226,827 tweets for the Bottom group. Content analysis refrains from correcting user-provided language (Zaucha & Agur, 2022). Latent Dirichlet Allocation modeling (Blei & Jordan, 2003) is employed to identify the topics in the two groups . Given the influence of the number of topics on the LDA model, an initial automated topic modeling is conducted 10 times for each group. Based on the overlap of topics, the optimal number of topics is determined to be 5, with typically only a small overlap between the last two topics, indicating a more complete and distinct categorization. After 30 runs with 5 topics for each group, the optimal topic distribution is derived. Finally, this study also extracts emoji types and quantities from the content in both groups and conducts a Mann-Whitney U for comparison.

## 5. Results

*5.1 Structural Features of the Network*

From a global structural perspective, the average degree is 2.733, and the degree distribution follows a power-law pattern. After applying the OpenORD layout, the entire retweet network is illustrated in figure 1. As depicted, a considerable number of highly connected nodes form a central cluster, while numerous loosely connected nodes are dispersed in the outer layers. The network comprises 37,541 bidirectional edges with unidirectional connections prevailing. The edge direction mirrors the flow of information in the network. Consequently, the network exhibits an asymmetric structure in terms of influence (Bulte & Joshi, 2007). In essence, most nodes are influenced by specific nodes but do not reciprocally influence these specific nodes. Instead, they continue to influence other nodes or have no impact at all.

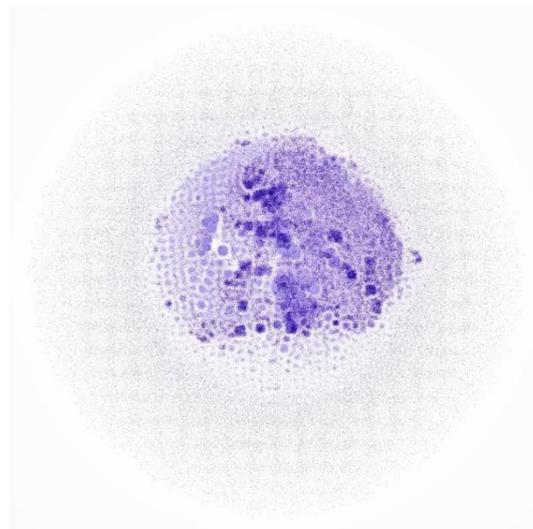

Fig. 1:Network topology based on OpenORD Layout. Nodes are colored and sized based on their degree, with darker and larger nodes representing higher degrees. Edges are not displayed in the graph.

*5.2 Structural Features of Core Communities*

To delve deeper into the structural features, it is imperative to pinpoint its core structure. Given diverse definitions across studies and acknowledging the network's large-scale complexity, we adopt 10 times the average degree of the network as a criterion for filtering the k-core (Goldenberg & Shapira, 2009). Subsequently, Modularity analysis is performed on this core network, yielding a modularity value of 0.648. Ultimately leads to the identification of 7 highly clustered communities, as illustrated in Figure 2.

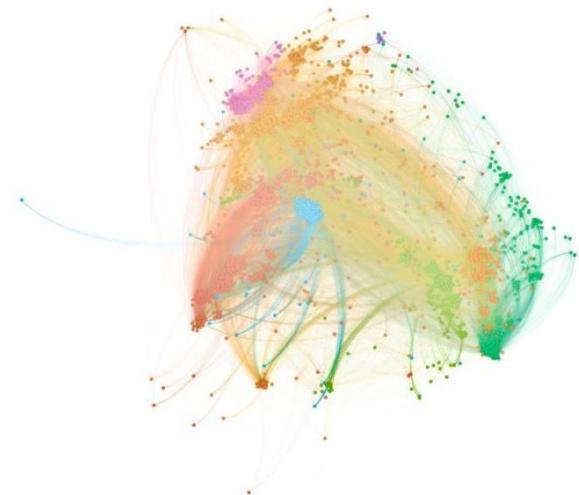

Fig. 2: The core network topology based on OpenORD Layout, comprises nodes with a degree of 27 or higher. This core network includes 10,766 nodes and 405,569 edges, with a mere 5,689 bidirectional edges.

*Community 0* is too small to offer significant insights into interaction patterns and is not considered for future analysis. *Community 1* has the largest size. *Community 5* has the highest number of bidirectional edges, indicating less pronounced asymmetry in influence direction and better interactivity. *Community 6* excels with the highest average clustering coefficient, implying strong connectivity. Conversely, *Communities 3 and 4* display higher average path lengths, suggesting sparser connections between nodes. Descriptive statistics for the six communities are outlined in Table 1.

| Community | Nodes | Edges (bidirectional edge) | Avg.degree | Avg.Clustering Coefficient | Avg.Path length |
|---|---|---|---|---|---|
| Community 1 | 4,756 | 186,776 (119) | 39.27 | 0.003 | 1.40 |
| Community 2 | 918 | 18,802 (73) | 20.48 | 0.006 | 2.90 |
| Community 3 | 1,012 | 37,288 (237) | 36.85 | 0.026 | 4.82 |
| Community 4 | 2,386 | 62,748 (819) | 36.85 | 0.021 | 4.22 |
| Community 5 | 733 | 31,672 (9734) | 43.21 | 0.238 | 2.36 |
| Community 6 | 954 | 31,743 (1187) | 33.27 | 0.096 | 2.53 |

Table 1: Descriptive statistics of structural characteristics of 6 communities

*5.3 Key Members and Behavioral Traits of Core Communities*

Next, attention is directed towards the internal dynamics of each community to abstract unique interaction patterns. The core structures of each community are depicted in Figure 3.

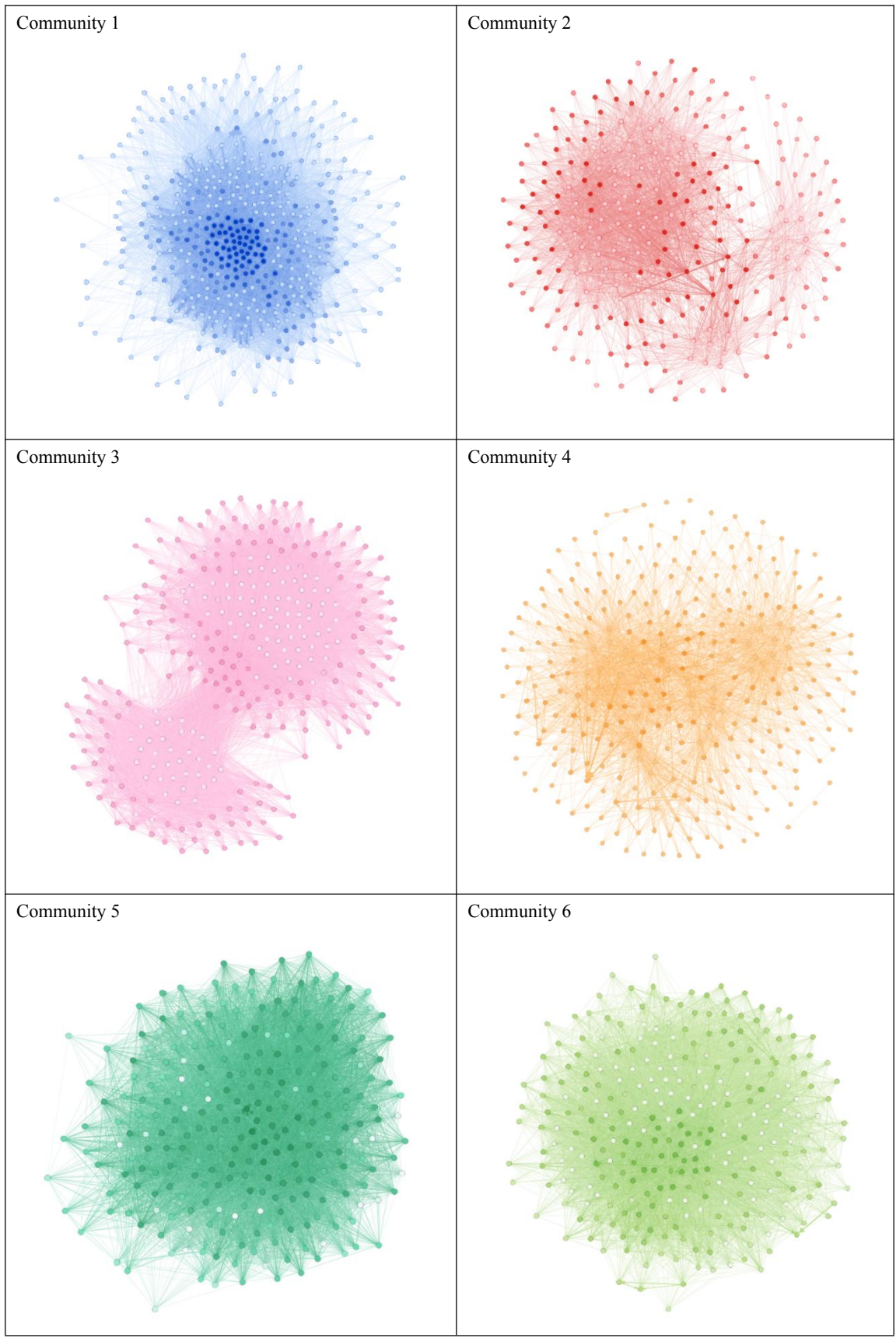

Fig. 3: A retweet network structure diagram based on the Fruchterman-Reingold layout. For clarity, the degree range is set to be twice the average degree of each community, with darker node colors indicating higher in-degrees.

The distribution of nodes within the community reveals distinct patterns. From a holistic view, these nodes either coalesce into cohesive clusters, as evident in *Communities 1, 5, and 6*, or can be subdivided into internal subgroups, illustrated by *Communities 2 and 3*. Upon scrutinizing the in-degree distribution of nodes across various positions, most communities demonstrate a bipolar nature. In one scenario, nodes with notably high in-degrees aggregate in the central core, while nodes with notably high out-degrees encircle the periphery, exemplified by *Communities 1* and *Communities 6*. Conversely, in an alternative scenario, such as *Community 3*, this pattern is reversed.

Bipolarity signifies that some nodes rarely retweet content from other nodes, with their content primarily being retweeted within the community, or vice versa—some nodes solely retweet content from other nodes, with their content seldom being retweeted. Only a very small fraction of nodes engage in both retweeting and being retweeted by other nodes.

Which nodes exhibit similar behavioral characteristics? Going deep into the context shaped by the intricate structure, the role composition within each community and the enduring patterns of interaction among them are obtianed, as shown in Figure 4.

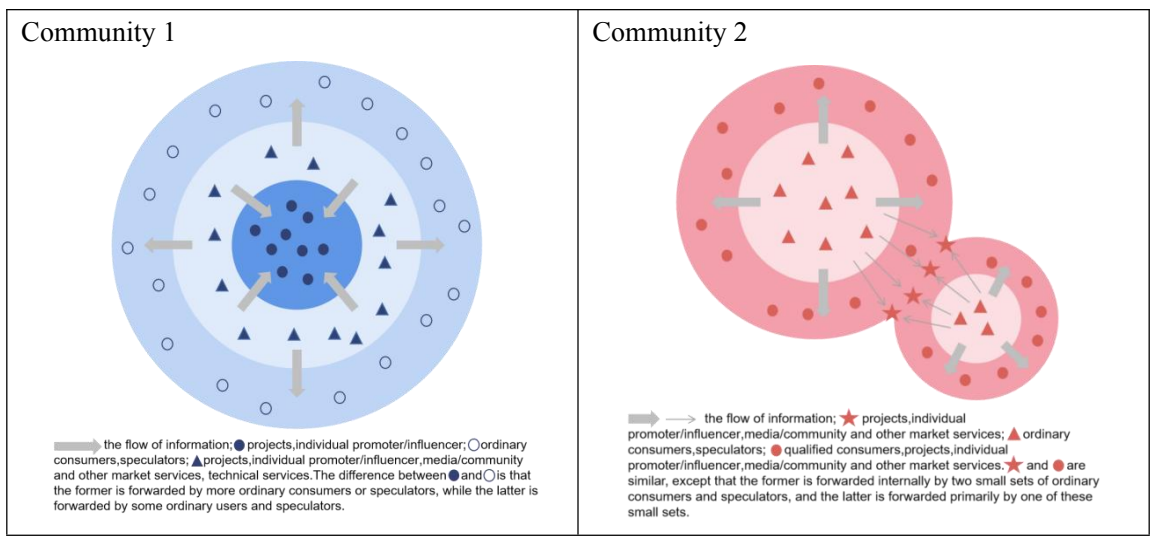

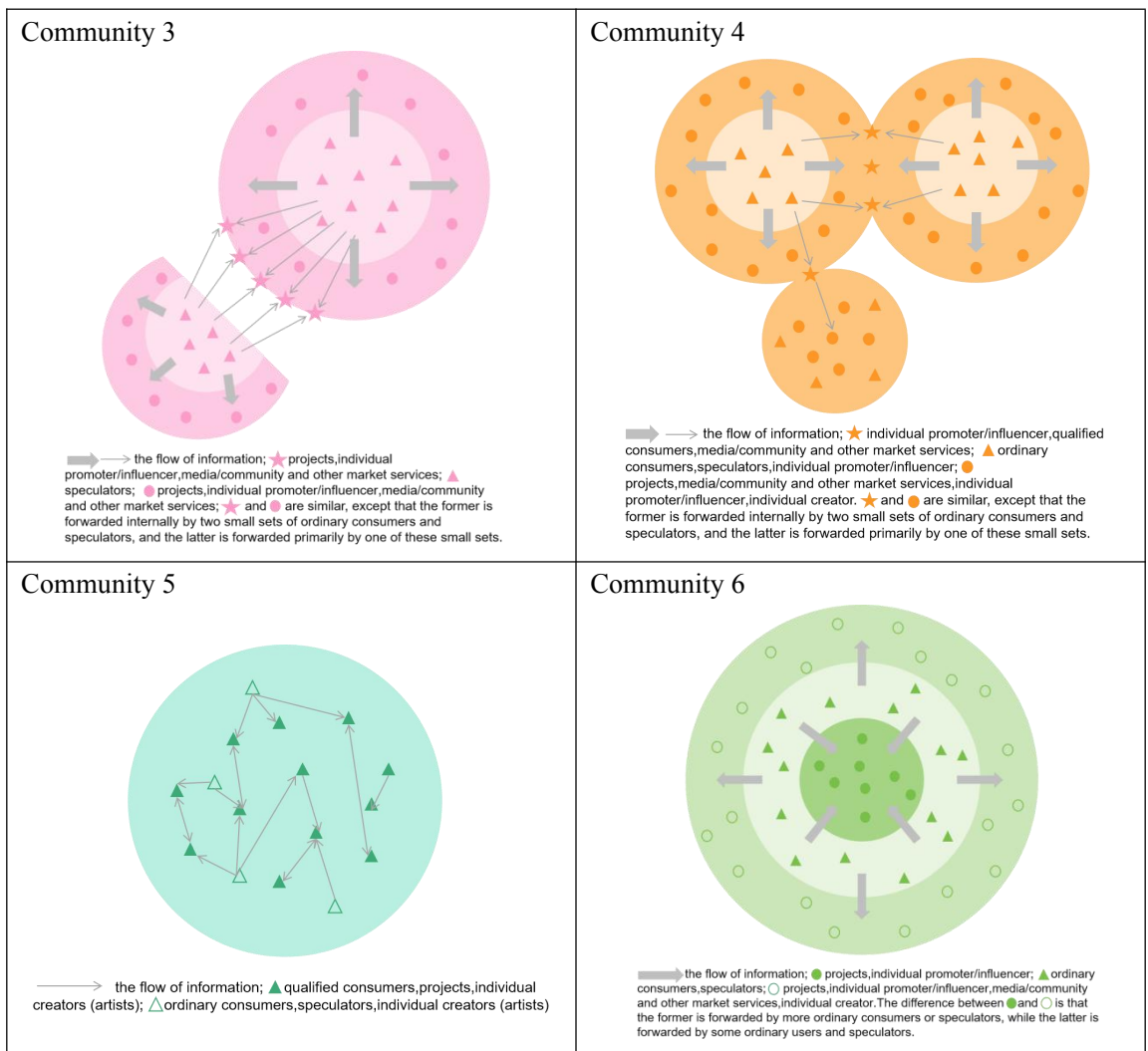

Fig. 4: A retweet network diagram for different communities, where different symbols represent different types of participants and the arrow direction is consistent with the in-degree.

The center of *Community 1* comprises nodes with high in-degrees, mainly consisting of projects, individual promoters/influencers, and qualified consumers. This community is chiefly influenced by organizational and individual promoters. The intermediate layer consists of nodes with high out-degrees and minimal in-degrees, largely consisting of ordinal consumers and speculators. The periphery includes nodes with relatively high in-degrees and similarly minimal out-degrees, mainly consisting of projects, individual promoters/influencers, while some from media/community and other market services, as well as technical services. Internal retweet relationships can be summarized as intermediate layer nodes retweeting almost all central nodes and partially retweeting peripheral nodes, forming a gear-like structure composed of multiple star networks.

*Community 6* closely resembles *Community 1*, with the distinction being that core users in *Community 6* are almost Japanese, and there are fewer speculators in the intermediate layer. Additionally, reciprocal retweet relationships exist among individual promoters/influencers, projects, and community volunteers in the periphery, indicating heightened interactivity.

In *Community 2*, two internal subgroups on the left and right are connected by a few bridge nodes, primarily consisting of media/community and other market services, individual promoters/influencers, as well as projects. The subgroups can be roughly divided into two parts. The first part consists of central nodes with high out-degrees and minimal in-degrees, primarily including ordinary consumers and speculators. The second part consists of peripheral nodes with high in-degrees and minimal out-degrees, mainly including qualified consumers, individual promoters/influencers, projects, or media/community and other market services. The relationships can be summarized as central nodes retweeting almost all peripheral nodes on the same side, and some central nodes also collectively retweet specific peripheral nodes. The left-side structure resembles a gear, differing in retweet direction from *Community 1*.

*Community 3* presents two clear internal subgroups. The upper-right one is associated with the Cardano blockchain, while the lower-left one is linked to the Cronos blockchain. Its structure is akin to *Community 2*, with more bridge nodes connecting the two internal subgroups.

*Community 4* presents an interior comprising three small subgroups connected by a few bridge nodes. The left and right subgroups connect through qualified consumers. Essentially, these qualified consumers share similarities with the peripheral nodes on both sides, characterized by high in-degrees and minimal out-degrees. The peripheral nodes mainly consist of media/community and other market services, individual promoters/influencers, projects, and individual creators/artists. The distinction lies in the fact that qualified consumers are retweeted by central ordinal consumers, less-followed individual promoters/influencers, and speculators on both sides, while peripheral nodes are only retweeted by central users on the same side. The structure of these two subgroups is similar to that of *Communities 2 and 3*. The left and lower-left subgroups are connected by a few individual promoters/influencers, where these nodes initially retweet media/community and other market services, individual promoters/influencers, and projects from the lower-left subgroup before being retweeted by central nodes on the left side.

*Community 5* features numerous nodes with high in-degrees and out-degrees, typically representing qualified consumers, projects that have been released and are still ongoing, and individual creators/artists. Mutual retweeting suggests heightened interactivity. Additionally, a small number of nodes with high out-degrees and minimal in-degrees consist of new individual creators/artists and ordinal consumers. The structure is challenging to abstract into a specific type, presenting a relatively mixed and diverse form.

*5.4 Topic Analysis of Different Influential Contents*

After performing LDA on the content of the Top group and Bottom group, the identified themes for each group are presented in Table 2. In essence, the most significant difference in themes between the Top and Bottom groups is the inclusion of Freemint within the former. Moreover, the content in the Top group tends to perceive NFTs as a type of crypto asset, whereas the Bottom group tends to view NFTs as a vehicle for web3 and the metaverse. The content in the Top group indicates a stronger emphasis on financial value.

| Top Group | | | |
|---|---|---|---|
| Topic | Rank | Count | Key Words & Example |
| Freemint Theme | 1 | 35,335 | nft, follow, drop, giveaway, nftgiveaway, eth, sell, nftcommunity, free, cnft |
| | | | airdrop excellent Japanese NFT!\|NFT Free Gallery\|\|✅Follow @oatmeal_mix_nft @kei31\|✅RT & Like\|✅Reply your Address\|⏰23h\|once to person\|\|The NFT Free Gallery will be transferred to the hidden area of your wallet |

| Issuance Theme | 2 | 122,761 | nft, crypto, eth, nftcommunity, nfts, opensea, openseaioassetsethereu, mint, new, collection |
| --- | --- | --- | --- |
| | | | 🎁Whitelist giveaway!🎁\|\|Another giveaway in discord. Join here:discord.gg/sM4hCgCEKZ\|and react in channel #giveaways 🎉\|Good luck!❤️ |
| Art Theme | 3 | 36,981 | nftart, nftcollector, digitalart, nftcollection, nfts, nftcommunity, cryptoart, nft, eth, art |
| | | | The Starry Night by Vincent Van Gogh 🎨\|\|Art history's most famous celestial scene and one of the world's most recognisable paintings 🌙⚫\|\|Three officially licensed digital collectibles drop on ElmonX Sun, 26 March 9 AM PT. |
| Blockchain Theme | 4 | 37,474 | nftcommunity, nft, nfts, theta, tezos, nftcollector, gm, buy, metaverse, tdrop |
| | | | #tezosArts #TezQuakeAid\|ARABESQUE\|19 for sale\|1.00 Tezos per piece\|The proceeds benefit the earthquake victims in Turkey and Syria |
| Memecoin | 5 | 22,715 | nft, quack, richquack, xrpl, twitfi, xrpname, lunr, sell, twitdao, de |
| | | | Top Memecoins With Highest Social Engagements\|\|$DOGE @dogecoin\|$FLOKI @realflokiinu\|$SHIB @Shibtoken\|$BABYDOGE @BabyDogeCoin\|$QUACK @richquack\|$VOLT @VoltInuOfficial\|$KATZ @katz_community\|$CATE @catecoin\|$TSUKA @dejitaru_tsuka\|$AKITA @AKITA_network |
| Bottom Group | | | |
| Topic | Rank | Count | Key Words & Example |
| Art Theme | 1 | 47,768 | opensea, nftcommunity, nftart, nft, nftartist, nfts, nftcollector, art, nftdrop, nftcollection |
| | | | hello I just released my latest digital art work on tezos nft, which is still part of the love and hope series\|\|'Fly in the air'\|JPEG\|1818 x 1818 pixels \|5 edition\|1 #xtz\| |
| Virtual Applications Theme | 2 | 87,387 | nft, nftcommunity, nfts, art, crypto, web, nftart, metaverse, project, collection |
| | | | Found the next WEB3 project I am into. 🔥 With the creator having more ⬆️ passion than I've seen in any other🤩, and the network that is already built along with its ideas and roadmap 🗺️ being more unique! STAND BY!! 📘🚀🚀🚀 |
| Crypto Theme | 3 | 46,604 | gpt, airdrop, crypto, nft, btc, eth, bitcoin, web, quack, hodl |
| | | | I have no doubt #GPT-4 is the next 10x coin! HODL tight and enjoy the ride! 🚀🔥 |
| Trading Theme | 4 | 23,771 | twitdao, nft, twitfi, de, bought, web, nfts, openseaioassetsethereu, mrtweet, eth |
| | | | '٦٢٠.eth ⚠️ bought for 0.289 ETH ($492.18) on OpenSea 🌊opensea.io/assets/ethereu…\| |
| Security Theme | 5 | 21,297 | wiboycoinecplaza, sgtkcoin, metaycoin, aexcoin, fastbitra, aax, expmarks, fund, holbilt |
| | | | Drop your complaint Now if you are unable to access your account in any of these platforms? |

Table 2: Examples of typical content for different groups

The *'Freemint Theme'* in the Top group mainly includes tweets related to freemint events, guidelines on selling NFTs, NFT trading news, and prompting transactions. Typically, freemint-related tweets consist of two key elements. First, information about rewards indicates the opportunity to acquire an NFT for free. Second, behavioral directives, often require users to follow, retweet, like, tag friends, and provide their wallet address, among other steps, within a specified time range.

Within the *'Issuance Theme'*, although it also includes a considerable amount of minting and selling content, it typically refers to new series, products, and phases of campaigns. Posters usually prompt users to take actions beyond retweeting, such as trading according to specific requirements or participating in other online community activities. Additionally, this theme often portrays NFTs as a subdomain within the crypto and involves information about industry developments. A noteworthy keyword in this topic is 'openseaioassetsethereu', originally a link to a trading marketplace, which disappears in the process of quote retweeting due to text copying.

The *'Art Theme'* of the Top group primarily revolves around the issuance and sale of artistic NFTs, along with promotion by individual artists. The *'Blockchain Theme'* is associated with the Theta and Tezos blockchains. An interesting keyword is 'gm', an abbreviation for 'good morning', with no special meaning but contributes a formalized characteristic to communication. The content of the *'Memecoin Theme'* is less related to NFTs and mainly focuses on meme tokens.

In the *'Art Theme'* of the Bottom group, there is an overview of the issuance, listing, and trading of a specific artistic NFT. This topic is always posted by individual creators, and predominantly involves OpenSea's personal trading marketplace.

The *'Virtual Applications Theme'* frequently mentions web3 and the metaverse. When referencing web3, similar to the *'Issuance Theme'* in the Top group, NFTs are regarded as a part of the web3 sector. When regarding the metaverse, akin to the *'Blockchain Theme'* in the Top group, NFTs are considered as a tool, indicating that holding certain NFTs is necessary to enter the metaverse, emphasizing the virtual utility of NFTs.

The *'Crypto Theme'* keywords all point to meme tokens, having little direct relation to NFTs. The term 'hodl' is an interesting jargon that signifies holding onto tokens steadfastly.

The *'Trading Theme'* addresses the tweets of NFT transactions. Unlike the *'Art Theme'* in the Top group, this topic focuses on traders and highlights who purchased an NFT at what price rather than detailing the sale of an NFT at a specific price. The term 'openseaioassetsethereu' also appears here, following a similar pattern as mentioned before. Some content is related to meme tokens and projects.

The *'Security Theme'* has minimal relevance to NFTs but rather pertains to fraud and security within the crypto sector. Most keywords are associated with meme coins or the names of exchanges. The dissemination of this content primarily serves as a cautionary and mutual assistance measure.

*5.5 Presentation Forms Analysis of Different Influential Contents*

Beyond thematic differences, we extracted 822 emojis from the Top group with a total count of 231,494, and 856 emojis from the Bottom group with a total count of 168,219. Generally, the Top group used a larger number of emojis, but the Bottom group exhibited a greater variety of emoji types. Many emojis had minimal occurrences, displaying a distinct long-tail distribution. The top 30 emojis in terms of frequency for both groups are presented in Table 3, with emoji meanings sourced from the EMOJIALL (EMOJIALL, n.d.).

| Rank | Group | Emoji | Meaning | Count | Group | Emoji | Meaning | Count |
|---|---|---|---|---|---|---|---|---|
| 1 | Top | 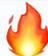 | fire | 25798 | Bottom | 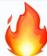 | fire | 16976 |
| 2 | Top | 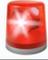 | police car light | 15666 | Bottom | 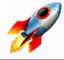 | rocket | 9835 |
| 3 | Top | 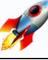 | rocket | 9911 | Bottom | 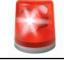 | police car light | 7361 |
| 4 | Top | 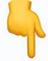 | backhand index pointing down | 8239 | Bottom | 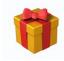 | wrapped gift | 6691 |
| 5 | Top | 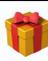 | wrapped gift | 7072 | Bottom | 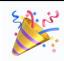 | party popper | 6392 |
| 6 | Top | 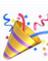 | party popper | 6057 | Bottom | 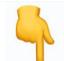 | backhand index pointing down | 5966 |
| 7 | Top | 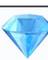 | gem stone | 5717 | Bottom | 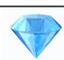 | gem stone | 4244 |
| 8 | Top | 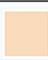 | light skin tone | 5078 | Bottom | 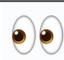 | eyes | 3472 |
| 9 | Top | 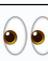 | eyes | 4865 | Bottom | 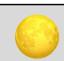 | full moon | 3369 |
| 10 | Top | 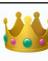 | crown | 4500 | Bottom | 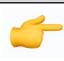 | backhand index pointing right | 3014 |
| 11 | Top | 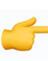 | backhand index pointing right | 3584 | Bottom | 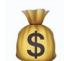 | money bag | 2994 |
| 12 | Top | 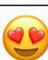 | smiling face with heart-eyes | 3569 | Bottom | 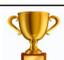 | trophy | 2918 |
| 13 | Top | 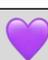 | purple heart | 3269 | Bottom | 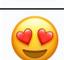 | smiling face with heart-eyes | 2427 |
| 14 | Top | 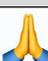 | folded hands | 3014 | Bottom | 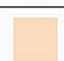 | light skin tone | 2409 |
| 15 | Top | 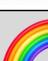 | rainbow | 2855 | Bottom | 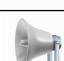 | loudspeaker | 2237 |
| 16 | Top | 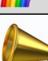 | megaphone | 2503 | Bottom | 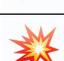 | collision | 2020 |
| 17 | Top | 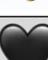 | black heart | 2126 | Bottom | 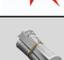 | rolled-up newspaper | 1732 |
| 18 | Top | 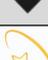 | dizzy | 2050 | Bottom | 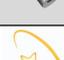 | dizzy | 1690 |
| 19 | Top | 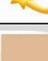 | medium-light skin tone | 2041 | Bottom | 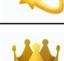 | crown | 1602 |
| 20 | Top | 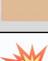 | collision | 1886 | Bottom | 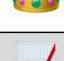 | chart increasing | 1301 |
| 21 | Top | 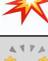 | raising hands | 1819 | Bottom | 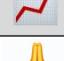 | folded hands | 1294 |
| 22 | Top | 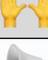 | loudspeaker | 1791 | Bottom | 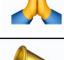 | megaphone | 1286 |
| 23 | Top | 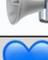 | blue heart | 1785 | Bottom | 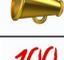 | hundred points | 1266 |
| 24 | Top | 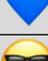 | smiling face with sunglasses | 1711 | Bottom | 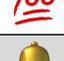 | bell | 1263 |
| 25 | Top | 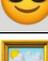 | framed picture | 1602 | Bottom | 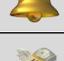 | money with wings | 1173 |
| 26 | Top | 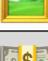 | dollar banknote | 1580 | Bottom | 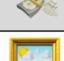 | framed picture | 1139 |
| 27 | Top | 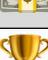 | trophy | 1546 | Bottom | 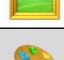 | artist palette | 1136 |
| 28 | Top | 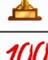 | hundred points | 1503 | Bottom | 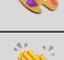 | clapping hands | 1133 |

| 29 | Top | 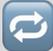 | repeat button | 1436 | Bottom | 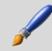 | paintbrush | 1133 |
|---|---|---|---|---|---|---|---|---|
| 30 | Top | 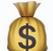 | money bag | 1431 | Bottom | 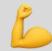 | flexed biceps | 1117 |

Table 3: An overview of the frequency of different emojis in different groups

Analyzing the top 30 emojis in frequency from both groups, 21 emojis are common to both, with 'fire' being the most prevalent in both groups. In practical usage, 'fire,' as well as 'police car light,' 'wrapped gift,' and 'party popper,' mostly appeared in freemint promotional content, emphasizing the theme of free giveaways. Other frequently used emojis in this theme include 'trophy,' various colored hearts, 'loudspeaker,' and 'collision.'

Other emojis are shared between the two groups. For example, 'rocket' symbolizes price increases or product launches. 'Gem stone' is often used to denote a valuable NFT. 'Backhand index pointing down' commonly indicates links or prompts comments. 'Eyes' and 'loudspeaker' generally emphasize specific information. 'Smiling face with heart-eyes' conveys feelings of liking and happiness, while 'crown' denotes characteristics or best-selling of NFTs. These emojis appear across various themes, displaying versatile usage across different contexts.

Furthermore, 9 emoji within the top 30 in frequency differ between the two groups. Notably, the 'repeat button,' prevalent in the Top group, is typically used to prompt users to retweet. It serves as an action-oriented emoji and is less frequently used in the Bottom group. Those representing price information and changes are exclusive to the Bottom group. For instance, 'full moon' is often paired with 'rocket,' symbolizing the slang 'to the moon,' expressing the hope for prices to continually rise to their peak. Table 4 outlines typical discourses associated with some emojis.

| Type | Emoji | Typical Discourse |
|---|---|---|
| High-frequency emojis appearing in both groups | 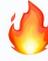 | 🔥INSANE GIVEAWAY🔥\|\|🎉3 HUGE Prizes + Instant Winners🎉\|\| 🦒#6452\|🦒#3332 (Zombie ranked #239 in Rarity)\|500 $GARY\|\|Follow \|@GaryLHenderson\|\|@giraffetowernft\|+ \|@GaryCoinOnSol \| \|✅Like & RT\|✅Tag 3 Friends\|\|Ends in 48 hours ⏰ \| |
| | | 🚨 GIVEAWAY CURRENTLY LIVE 🚨\|\|1x @alienfrens #3607\|\|📣 GET YOUR ENTRIES IN NOW!! 📣\|\| 🚨 @nftkoosh x @kooshgivesback 🚨\| |
| | | 🎁 GIVEAWAY 🎁\|\| Prizes: \|\|🏆 $30 in $ETH\|\| Entry: \|1️⃣ Follow @BAIBERC20 \|2️⃣ Like + RT \|3️⃣ Tag friends \|\|⏰ 24hr ⏰\| |
| | 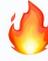 | 🎉1 ethereum price!!\|\|📈 Get your tickets on opensea🚀\|\|🤝1 Ticket = 2 Matic\|\|🔥Only 1000 tickets \|\|☆1 winner☆ |
| | 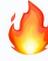 | Look at it!😍 This is a new Gem!💎 Every 2 weeks they conduct sweepstakes!🙂\|Beta is launched!😎 |
| | 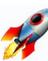 | 3 words that describe your NFT project 👇👇👇 |
| | 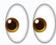 | Hey there 🙌👀 ‖Wanna mint a @PeculiarPugs #NFT for 0.025 ETH THATS A 75% DISCOUNT Y'ALL! ‖Tell me your favorite things about Pugs and the PeculiarPugs to participate ❤️😀 |
| Exclusive high-frequency emojis in the Top group | 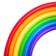 | Gm! 🌈💜🐻☕🙌✨‖Say it back so I can follow the you 💜😀 |
| | 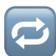 | 📢Minting Maniac is back!📢‖Every day until the release of the shimmerEVM testnet we will #giveaway a #NFT of the IotaOrigin genesis collection. EVERY DAY! #IOTA #SMR #Shimmer‖$AUREUS ‖💙 Like✅ Follow🔁 Retweet🎟 get a whitelist spot on our discord! |
| | 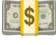 | Drop your #NFT 👇 For Sales 💰<br><br>🚨 DEGEN #1404/3221 SOLD!!! => Price 0, 09 $SOL ($2.36 💸) ‖🔻RANK M: N/R 🔻RANK HR: N/R ‖🖌FP: 0, 09 $SOL 👥Holders: 893 🚀Price Mint: 0, 059 $SOL‖SATTONFT MINT NOW 🛒 |
| Exclusive high-frequency emojis in the Bottom group | 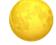 | what nft projects are going to the moon? 🚀🌕🔥 |
| | 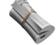 | 'It's official! 🔊‖Take a guess and you could win an FS Key NFT. 🔥‖Enable notifications and stay tuned for the latest news. 🤩 |
| | 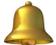 | $SOL Giveaway ‖🏆1 Winner $SOL‖1️⃣ Follow @MonkeyNightCLUB ‖2️⃣ RT like，turn on notifications🔔🔔🔔‖⏰180 min ‖ |

Table 4: Examples of typical discourses for two groups

Beyond emoji categories, this study compared the differences in the quantity of emoji usage between the Top group and Bottom group. After excluding category-specific emojis unique to each group, 730 common emojis are retained. The Mann-Whitney U test is conducted due to the extreme left-skewed distribution of emoji usage in both groups. The results shows a significant difference in emoji usage counts between the two groups (p = 0.0211 < 0.05), with the Top group generally utilizing a greater number of emojis compared to the Bottom group.

## 6. Discussion

### 6.1 Three Different Information Mechanisms

According to the network structure, three distinct information mechanisms can be identified. The first, exemplified by *Community 1 and 6*, manifests as a 'Ringed-layered' pattern primarily comprising projects, individual promoters/influencers, and speculators. In this pattern, information flow is typically unidirectional. There exists a conspicuous asymmetry in influence, with centrally positioned projects and individual promoters/influencers having their content retweeted by numerous peripheral speculators. However, the reverse does not work. Interactions between projects, individual promoters/influencers, and speculators remain limited. The structural configuration mirrors a gear, constituted by an amalgamation of numerous star networks.

The second, represented by *Community 2, 3, and 4*, presents a 'Group-associative' pattern consisting of several communities. The center nodes of each community may exhibit either high in-degree or high out-degree, while the periphery is the opposite. The structure of a single community mirrors the previously mentioned gear-like shape. Different communities establish interconnections through bridge nodes such as project teams, individual promoters, and intermediaries like media/community market services.

The third, epitomized by *Community 5*, embodies an 'Interactive' pattern mainly involving qualified consumers and individual creators/artists. The complexity of participant interactions in this pattern makes it challenging to delineate a specific structure.

The prevalence of 'Ringed-layered' and 'Group-associative' pattern suggests that marketers aspiring to trigger viral spread accompanied by economic incentives in social networks must acknowledge the presence and significance of broadcasting spread. Viral spread can sometimes be an idealistic wish, and broadcasting spread may facilitate to building of institutional relationships, thereby promoting diffusion. Here, the institution refers to reciprocity, wherein users receive economic incentives, and marketers achieve their marketing goals.

*6.2 Purposefulness and Potential Organizational Tendencies of Behavior*

Why do Participants connect in specific ways? Text minging reveals that these Participants gathering due to economical incentives, which are deeply embedded in the social network. In the content posted by high-retweet nodes, the most weighted and extensive topics are related to the minting of NFTs, particularly associated with the initial freemint stage. Moreover, these contents tend to perceive NFTs as a new form of digital asset.

Freemint is a unique way for NFTs issuance, centered around the anticipation of obtaining profits without cost. Participants, regardless of whether they are qualified consumers, ordinal consumers, or speculators, can acquire an NFT at nearly zero cost by engaging in specified actions, including retweeting, within a defined time range in the social network. Subsequently, these NFTs can be traded for profit in the secondary market. This factor significantly contributes to the dominance of the 'Ringed-layered' pattern in the diffusion of NFTs.

Speculators achieve maximum returns through indiscriminate retweeting of freemint-related content. Given that speculators are not required to communicate with their peers, thus influence within the network often exhibits asymmetry. Participants also engage in retweeting activities related to the listing and sale of artistic NFTs or trends in the cryptocurrency market, with the common objective of enhancing economic incentives. Furthermore, some participants retweet content associated with 'Security Themes' as a means of warning or mutual support.

In addition to the purposefulness of behavior, participants also show organizational tendencies. Some speculators use the same background image, and a few even use the term 'crew' on their profiles, possibly indicating that speculators may have their community to mass-forward messages upon hearing about freemint.

*6.3 The Catalyzing Influence of High-Frequency Posting and Emoticon Usage*

Examining the quantity of content posted by nodes with varying in-degrees, the top 25% in-degree nodes averaged 15.16 posts, while the bottom 25% in-degree nodes averaged 4.49 posts. The observed advantage in retweeting among nodes with high in-degrees aligns with their heightened activity levels, consistent with previous findings (Iyengar et al., 2011). Additionally, the observation that nodes in the top 25% in terms of in-degree use a greater number of emojis.

The finding that increased emoji usage may facilitate user forwarding behavior is consistent with previous results (McShane et al., 2021).

## 7. Conclusion

This paper delineates the diffusion of digital innovation in social networks under an economic context by analyzing retweet networks and NFT-related content on Twitter. Three information mechanisms are abstracted during the diffusion process, and future research can delve deeper into the functionalities of these structure features. Additionally, the paper tentatively explores the purpose of behavior by examining the limited discourse consciousness. Recognizing the non-equivalence of discourse and practical consciousness, future studies can employ qualitative methods such as interviews and participatory observations to further explore the motivations behind retweeting.

The scalability of this study lies in the gear-like network structure composed of numerous star networks, offering an experiential framework for digital marketing practices in social networks. As more marketers aim for viral spread through economic incentives like coupons and discounts, reliable simulations face challenges in grasping diffusion microstructures. More research could also discuss the dynamic relationship between rules and structure, as well as between speculators and target users.